\documentclass[page-classic]{epl2} % for 2 columns style with line numbers
\usepackage{amsmath,amssymb,amscd}
\usepackage{graphicx}
\usepackage{bm}
\usepackage{color}

\usepackage{subfigure}

\graphicspath{{images/}{plots/}}

\newcommand{\de}{\mathrm{d}}
\newcommand{\e}{\mathrm{e}}

\newcommand{\bea}{\begin{eqnarray}}
\newcommand{\eea}{\end{eqnarray}}
\newcommand{\veps}{\varepsilon}

\renewcommand{\vec}[1]{\bm{#1}}

\newcommand{\ie}{\textit{i.e.\ }}
\newcommand{\eg}{\textit{e.g.\ }}

\newcommand{\eq}{eq.~}

\newcommand{\fig}{fig.~}

\title{Ground state energy of noninteracting fermions with a random energy spectrum}
\shorttitle{} %Insert here a short version of the title if it exceeds 70 characters

\author{Hendrik Schawe \inst{1} \and Alexander K. Hartmann \inst{1} \and Satya N. Majumdar \inst{2} \and Gr\'egory Schehr \inst{2}  }
\shortauthor{}

\institute{
\inst{1} Institut f\"ur Physik, Universit\"at Oldenburg, 26111 Oldenburg, Germany \\
  \inst{2} Univ. Paris-Sud, CNRS, LPTMS, UMR 8626, Orsay F-91405, France
}


\abstract{We derive analytically the full distribution of the
ground-state energy of $K$ non-interacting fermions in a disordered
environment, modelled by a Hamiltonian whose spectrum consists of $N$
i.i.d.~random energy levels with distribution $p(\veps)$ (with $\veps
\geq 0$), in the same spirit as the ``Random Energy Model''. We show
that for each fixed $K$, the distribution $P_{K,N}(E_0)$ of the
ground-state energy $E_0$ has a universal scaling form in the limit of
large $N$. We compute this universal scaling function and show that it
depends only on $K$ and the exponent $\alpha$ characterizing the small
$\veps$ behaviour of $p(\veps) \sim \veps^\alpha$.
We compared the analytical predictions with results from numerical simulations.
For this purpose we employed a sophisticated
importance-sampling algorithm that allowed us to obtain
the distributions over a large range of the support down to
probabilities as small as  $10^{-160}$. We found asymptotically a very
good agreement between analytical predictions and numerical results.
}

\begin{document}

\maketitle

The celebrated ``Random Energy Model'' (REM) of Derrida \cite{REM1} has continued to play a central
role in understanding different aspects of classical disordered systems, including spin-glasses,
directed polymers in random media and many other systems. In the REM, one typically has
$N$ energy levels which are considered to be independent and identically distributed (i.i.d.)
random variables, each drawn from a probability distribution function (PDF) $p(\varepsilon)$. Typical observables
of interest are the partition function, free energy, etc. The REM can also be useful as a toy model
in quantum disordered systems. For example, let us consider a single quantum particle in a disordered
medium with the Hamiltonian $\hat h$. We will assume that the spectrum of the operator $\hat h$ has a finite
number of states $N$ (for instance a quantum particle on a lattice of finite size and a random onsite potential,
as in the Anderson model). In general, solving exactly the spectrum of such an operator is hard, for a generic random
potential. One possible approximation, in the spirit of the REM in classical disordered systems, would be to
consider the toy model where one replaces the spectrum of the actual Hamiltonian by $N$ {\it ordered}
i.i.d.~energy levels $\varepsilon_1 \leq \varepsilon_2 \leq \cdots \leq \varepsilon_N$ each drawn from the common
PDF $p(\varepsilon)$. Without loss of generality,
we will also assume that the Hamiltonian $\hat h$ has only positive eigenvalues. This would mean that, in the
corresponding toy model, the PDF $p(\varepsilon)$ is supported on $[0, +\infty)$. It is well known that, in a strongly disordered quantum system, where all single-particle eigenfunctions are localised in space, the energy levels can be approximated by i.i.d.~random variables (see \eg \cite{MNS}). Therefore the REM that we consider here will be relevant in such strongly localised part of the spectrum of a disordered Hamiltonian.

Now consider a system of $K$ noninteracting fermions with the
Hamiltonian $\hat H_K = \sum_{i=1}^K \hat h_i$ where $\hat h_i$ is the
single particle Hamiltonian associated with the $i$-th particle. The
ground state of this many-body system would correspond to filling up
the single particle spectrum up to the Fermi level $\veps_K$, with one
particle occupying each of the states with energies $\veps_1, \veps_2,
\cdots, \veps_K$. The ground state energy $E_0$ of this many-body
system is therefore given by
\bea\label{def_E0}
E_0 = \sum_{i=1}^K \veps_i \;.
\eea
Clearly, $E_0$ is a random variable, which
fluctuates from one realisation of the disorder to another. Given
$p(\veps)$, we are interested in computing the distribution $P_{K,N}$
of $E_0$, for fixed $K$ (\ie the number of fermions) and $N$ (\ie the
number of levels). We note that, for $K=1$,
$E_0 = \veps_1$ is just the minimum of a set of $N$ i.i.d.~random variables
and is described by the well-known extreme value statistics \cite{Gumbel}.
Thus, for general value of $K$, in particular, it would be interesting to know
how sensitive the distribution of $E_0$ is to the choice of
$p(\veps)$. For instance, is there any universal feature of the
distribution of $E_0$ that is independent of $p(\veps)$? We note that
$E_0$ is a sum of random variables, but these random variables are not
independent due to the ordering $\veps_1 \leq \veps_2 \leq \cdots \leq
\veps_N$ (even though the original unordered random variables are
independent). Had they been independent, the sum $E_0$ in
\eq(\ref{def_E0}), by virtue of the Central Limit Theorem, would
converge to a shifted and scaled Gaussian random variable. Here, this
is not the case, as the ordering induces non-trivial correlations
between these variables.

In this paper, we compute exactly the PDF $P_{K,N}(E_0)$ for arbitrary $K$, $N$ and $p(\veps)$ and show that, indeed, a universal
feature emerges in the large $N$ limit. It turns out that the limiting distribution of $E_0$, for large $N$, depends only on the small $\veps$ behaviour of
$p(\veps) \approx B\, \veps^\alpha$, with $\alpha > -1$, but is otherwise independent of the rest of the features of $p(\veps)$.
For fixed $\alpha$ and fixed $K$, as $N \to \infty$, we show that the distribution of the ground state energy converges to a limiting scaling form
\bea\label{main_result_form}
P_{K,N}(E_0) \approx b\,N^{\frac{1}{\alpha +1}}  F_K^{(\alpha)} \left(b\,N^{\frac{1}{\alpha +1}}\, E_0 \right)
\eea
where $b = (B/(\alpha+1))^{1/(\alpha+1)}$ is just a scale factor. The scaling function $F_K^{(\alpha)}(z)$ (with $z \in [0,+\infty)$) is universal and depends only on $\alpha$ and $K$. We show that the Laplace transform of  $F_K^{(\alpha)}(z)$ is given explicitly by
\bea\label{expr_F_Laplace}
\int_0^\infty F_K^{(\alpha)}(z) \e^{-\lambda\, z} \, \de z= \frac{(\alpha+1)^K}{\Gamma(K) \lambda^{(\alpha+1)(K-1)}} \int_0^\infty x^\alpha \e^{-\lambda \,x - x^{\alpha+1}} \left[\gamma(\alpha+1,\lambda \, x)\right]^{K-1}\, \de x \;,
\eea
where $\gamma(a,x) = \int_0^x \de u \, u^{a-1} \e^{-u}$ is the incomplete gamma function. While we can invert formally this Laplace transform (\ref{expr_F_Laplace}), it does not have a simple expression for generic $\alpha$. However, we can derive the asymptotic behaviour of $F_K^{(\alpha)}(z)$
\bea\label{F_asymptotics}
F_K^{(\alpha)}(z) \approx
\begin{cases}
& c_1 \, z^{(\alpha+1)K-1} \;, \; \hspace*{1.9cm} z \to  0 \;, \\
& \\
& c_2 \, z^\alpha \, \exp{\left[-\left(\dfrac{z}{K}\right)^{\alpha+1}\right]} \;, \;\;\;\;z \to \infty\;,
\end{cases}
\eea
where $c_1 = \frac{[\Gamma(\alpha+2)]^K}{\Gamma(K+1) \Gamma((\alpha+1)K)}$ and $c_2 = \frac{(\alpha+1)K^{K-\alpha-2}}{\Gamma(K)}$ are constants. For
the extreme-value case $K=1$
our result, $F_1^{(\alpha)}(z) = (\alpha+1)\,z^\alpha \, \e^{-z^{\alpha+1}}$, coincides with the well known Weibull scaling function  \cite{Gumbel}. Note that here we are interested in the sum of $K$ lowest i.i.d.~variables supported over $[0,+\infty)$. We remark that in the statistics literature, in a completely different context, the sum of the top $K$ values of a set of i.i.d.~random variables with an unbounded support has been studied \cite{Nag1, Nag2}. However, we have not found our results (\ref{main_result_form}) and (\ref{expr_F_Laplace}) in the statistics literature.

We start with a set of $N$ positive i.i.d.~random variables $\{x_1, x_2, \cdots, x_N \}$, each drawn from a common distribution $p(x)$, supported on $[0, +\infty)$. The joint distribution of these variables is simply $P(x_1, \cdots, x_N) = \prod_{i=1}^N p(x_i)$. At this stage, these variables are unordered. We are interested in the first $K$ ordered variables $\{\veps_1, \veps_2, \cdots, \veps_K\}$ with $K\leq N$. This ordering makes these $K$ variables correlated. Indeed, the joint distribution of the $K$ lowest ordered variables can be written explicitly as
\bea\label{joint_ordered}
P(\varepsilon_1, \cdots, \varepsilon_K) =\frac{\Gamma(N+1)}{\Gamma(N-K+1)}\, \prod_{i=1}^K p(\veps_i) \, \prod_{i=2}^K \Theta(\veps_i - \veps_{i-1}) \, \left[\int_{\veps_K}^\infty p(u) \, \de u \right]^{N-K} \;.
\eea
This result can be easily understood as follows. We first choose the $K$ distinct variables from an i.i.d.~set of $N$ variables. The
number of ways this can be done is simply the combinatorial factor $N(N-1)\cdots (N-K+1)= \Gamma(N+1)/\Gamma(N-K+1)$ in \eq(\ref{joint_ordered}). The probability that they are ordered is $\prod_{i=1}^K p(\veps_i) \, \prod_{i=2}^K \Theta(\veps_i - \veps_{i-1})$, where the Heaviside theta functions ensure the ordering. In addition, we have to ensure that the $N-K$ remaining variables are bigger than $\veps_K$, \ie the largest value among the first $K$ ordered variables. Since these $N-K$ variables are i.i.d., this gives the last factor in \eq(\ref{joint_ordered}). The formula in \eq(\ref{joint_ordered}) is exact for any $p(\veps)$, $K$ and $N$. Given the joint PDF (\ref{joint_ordered}), we are interested in the distribution $P_{K,N}(E_0)$ of the ground-state energy $E_0$ in \eq(\ref{def_E0}). We therefore have
\bea\label{PKN_1}
P_{K,N}(E_0) = \int P(\varepsilon_1, \cdots, \varepsilon_K) \delta\left( E_0 - \sum_{i=1}^K \varepsilon_i \right) \prod_{i=1}^K \de \veps_i \;.
\eea
The form of this equation naturally suggests to consider the Laplace transform with respect to (w.r.t.) $E_0$
\bea\label{Laplace1}
\langle \e^{-s E_0}\rangle = \int_0^\infty P_{K,N}(E_0)\, \e^{-s E_0} \, \de E_0 \;.
\eea
Taking the Laplace transform of \eq(\ref{PKN_1}) gives
\bea\label{Laplace2}
\langle \e^{-s E_0}\rangle = \frac{\Gamma(N+1)}{\Gamma(N-K+1)} \int_0^\infty \de \veps_K\, p(\veps_K) \e^{-s \,\veps_K} \left[\int_{\veps_K}^\infty p(u) \de u \right]^{N-K} \, J_{K-1}(\veps_K) \;,
\eea
where
\bea\label{def_JK}
J_{K-1}(\veps_K) = \int \prod_{i=1}^{K-1}  p(\veps_i) \e^{-s \veps_i} \, \de \veps_i \prod_{i=2}^{K} \Theta(\veps_i - \veps_{i-1}) \;.
\eea
This multiple integral (\ref{def_JK}) has a nested structure and can be evaluated easily by induction and one gets
\bea\label{JK_expl}
J_{K-1}(\veps_K) = \frac{1}{(K-1)!} \left[\int_0^{\veps_K} \de u \, \e^{-su} p(u) \right]^{K-1} \;.
\eea
Using this result in \eq(\ref{Laplace2}), and also replacing, for later convenience, $\int_y^\infty \de u\, p(u) = 1 - \int_0^y \de u \, p(u)$, we get the exact formula
\bea\label{Laplace3}
\langle e^{-s E_0}\rangle = K {N \choose K} \int_0^\infty \de y \,p(y)\, \e^{-sy} \left[1- \int_0^y \de u\, p(u)\right]^{N-K} \left[ \int_0^y \de v\, p(v)\, \e^{-sv} \right]^{K-1} \;.
\eea
This formula has a simple interpretation. Taking the Laplace transform is equivalent to breaking the system into two species of random variables of size $K$ and $N-K$ (this can be done in $N \choose K$ ways): Each member of the first species of size $K$ comes with an effective weight $p(\veps)\,\e^{-s\veps}$, while in the second species of size $N-K$ each member comes with an effective weight $p(\veps)$. We first fix the $K$-th variable to have a value $y$, whose weight is $p(y)\, \e^{-sy}$. The members of the second species should each be bigger than $y$ (explaining the factor $ \left[ \int_y^\infty p(u)\,\de u\right]^{N-K}$), while the rest of the $(K-1)$ members of the first species should each be smaller than $y$, explaining the factor $\left[ \int_0^y \de v\, \e^{-sv} p(v)\right]^{K-1}$. Finally, the biggest variable among the members of the first species can be any of the $K$ members, explaining the factor $K$ multiplying the binomial coefficient $N \choose K$ in \eq(\ref{Laplace3}). With this interpretation, it is clear that \eq(\ref{Laplace3}) can be easily generalized to any linear statistics of the form $L_K = \sum_{i=1}^K f(\veps_i)$ where $f(\veps)$ is an arbitrary function. The ground state energy $E_0$ considered here corresponds to choosing $f(\veps) = \veps$. For general $f(\veps)$ the effective weight of each member of the first species discussed above is just $p(\veps)\, \e^{-s f(\veps)}$. Hence the formula in \eq(\ref{Laplace3}) generalises to
\bea\label{Laplace4}
\langle \e^{-s L_K}\rangle = K {N \choose K} \int_0^\infty \de y \,p(y)\, \e^{-sf(y)} \left[ \int_y^\infty p(u)\,\de u\right]^{N-K} \left[ \int_0^y \de v\, p(v)\,\e^{-sf(v)} \right]^{K-1} \;.
\eea
In this paper, we will focus only on the case $f(\veps) = \veps$. Below, we thus start with the exact result in \eq(\ref{Laplace3}) and
analyse its behaviour in the large $N$ limit.

To understand the large $N$ scaling limit, it is instructive to start with the $K=1$ case. In this case, $E_0 = \veps_1$ is just the minimum of a set of $N$ i.i.d.~random variables, each drawn from $p(\veps)$. In this case, \eq(\ref{Laplace3}) reads (upon setting $K=1$)
\begin{eqnarray}\label{keq1}
\langle \e^{-sE_0}\rangle = N \int_0^\infty \de y \, p(y) \, \e^{-sy}\left[1-\int_0^y \de u\, p(u) \right]^{N-1} \;,
\end{eqnarray}
where we replaced $\int_y^\infty \de u \, p(u) = 1 - \int_0^y \de u\, p(u)$, using the normalisation of $p(u)$. In the large $N$ limit, the dominant contribution to the integral over $y$ comes from the regime of $y$ where the integral $\int_0^y \de u\, p(u)$ is of order $O(1/N)$. For other values of $y$, the contribution is exponentially small in $N$, for large $N$. Hence, we see that, in the large $N$ limit, only the small $y$ behaviour of $p(y)$ matters. Let
\begin{eqnarray}\label{p_y_small}
p(y) \underset{y\to 0}{\approx} B\, y^\alpha
\end{eqnarray}
where $\alpha > -1$ in order that $p(y)$ is normalisable and clearly $B>0$. Substituting this leading order behaviour of $p(y)$ for small $y$ (\ref{p_y_small}) in \eq(\ref{keq1}), we get
\begin{eqnarray}\label{keq1:1}
\langle \e^{-s E_0}\rangle \approx B\, N \int_0^\infty \de y\, y^{\alpha} \, \e^{-sy} \exp{\left(-\frac{B\,N}{\alpha+1}\, y^{\alpha+1}\right)} \;.
\end{eqnarray}
Performing the change of variable $y = \left(\frac{\alpha+1}{B \, N} \right)^{\frac{1}{\alpha+1}} \, x$, we get
\begin{eqnarray}\label{keq1:2}
\langle \e^{-s E_0}\rangle \approx (\alpha + 1) \int_0^\infty \de x \, x^{\alpha} \exp{\left(-\frac{s}{b\, N^{\frac{1}{\alpha+1}}}\,x - x^{\alpha+1}\right)} \;,
\end{eqnarray}
where $b = (B/(\alpha+1))^{1/(\alpha+1)}$. Inverting the Laplace transform formally, we obtain the scaling form given in \eq(\ref{main_result_form}) with $K=1$ and the scaling function $F_1^{(\alpha)}(z)$ has its Laplace transform as in
(\ref{expr_F_Laplace}) with $K=1$. Inverting this Laplace transform exactly, we recover the Weibull scaling function $F_1^{(\alpha)}(z) = (\alpha+1) z^\alpha \,\e^{-z^{\alpha+1}}$. The calculation for $K=1$ shows that only the small $y$ behaviour of $p(y)$ matters in the limit of large $N$. Furthermore, for $K=1$, we see that the typical value of $E_0$ scales as $N^{-\frac{1}{\alpha+1}}$ for large $N$. We then anticipate that, even for $K>1$, the typical scale of $E_0$ will remain the same $E_0 \sim N^{-\frac{1}{\alpha+1}}$ for large $N$. Below, we indeed use this typical scale for $E_0$ (and verify a posteriori) and compute the scaling function $F_K^{(\alpha)}(z)$ for general $K$ in \eq(\ref{main_result_form}).

We now derive the main results in Eqs. (\ref{main_result_form}) and (\ref{expr_F_Laplace}) for all $K \geq 1$. Anticipating the scaling $E_0 \sim N^{-\frac{1}{\alpha+1}}$ as mentioned above, we set
\begin{eqnarray}\label{scale_E0}
E_0 = \frac{1}{b} \,N^{-\frac{1}{\alpha+1}} \, z \;,
\end{eqnarray} where $b$ is a
constant to be fixed later and the scaled ground state energy $z$ is
of order~$O(1)$. Substituting this scaling form (\ref{scale_E0}) in
\eq(\ref{Laplace3}), we see that the left hand side (l.h.s.) reads,
$\langle \e^{-sE_0}\rangle = \langle \e^{- s \, N^{-\frac{1}{\alpha+1}}
\,z/b}\rangle = \langle \e^{-\lambda\,z}\rangle$ where $\lambda =
N^{-\frac{1}{\alpha+1}}\, s/b$ is the rescaled Laplace variable. We
will take the $N \to \infty$ limit, keeping $\lambda$ fixed. This then
corresponds to $s \to \infty$ limit. On the right hand side (r.h.s.)
of \eq(\ref{Laplace3}) we make a change of variable $s\,y = \tilde x$
as well as $u = \tilde u/s$ and $v = \tilde v/s$. This gives
\begin{equation}\label{asympt1}
\langle \e^{-\lambda z} \rangle =  \frac{K}{s^K} {N \choose K} \int_0^\infty \de \tilde x\, p\left( \frac{\tilde x}{s}\right) \e^{-\tilde x} \left(1 - \frac{1}{s}\int_0^{\tilde x} \de \tilde u\, p\left( \frac{\tilde u}{s}\right) \right)^{N-K} \hspace*{-0.2cm}\left(\int_0^{\tilde x} \de \tilde v\, p\left(\frac{\tilde v}{s}\right)\,\e^{-\tilde v} \right)^{K-1}.
\end{equation}
In the large $s$ limit, we use $p(y) \approx B\,y^{\alpha}$ to leading order. Inserting this behaviour in \eq(\ref{asympt1}), we get
\bea\label{asympt2}
\langle \e^{-\lambda z} \rangle \approx  \frac{K\, B^K}{s^{(\alpha+1)K}} {N \choose K} \int_0^\infty \de \tilde x\, \tilde x^\alpha\,\e^{-\tilde x} \left(\e^{-\frac{B(N-K)\,}{(\alpha+1)\,s^{\alpha+1}}\,\tilde x^{\alpha+1}  } \right) \left[\gamma(\alpha,\tilde x)\right]^{K-1}
\eea
where we recall that $\gamma(a,x) = \int_0^x \de u \, u^{a-1} \e^{-u}$ is the incomplete gamma function. We now use $s = (\lambda\,b)\, N^{\frac{1}{\alpha+1}}$ and choose
\begin{eqnarray}\label{eq:B0}
b = \left(\frac{B}{\alpha+1}\right)^{\frac{1}{\alpha+1}} \;.
\end{eqnarray}
Furthermore, in the large $N$ limit $K {N \choose K} \sim N^K/\Gamma(K)$. Using these results, and rescaling $\tilde x = \lambda \,x$, we arrive at
\bea\label{Laplace5}
\langle \e^{-\lambda\,z}\rangle  = \frac{(\alpha+1)^K}{\Gamma(K) \lambda^{(\alpha+1)(K-1)}} \int_0^\infty x^\alpha \e^{-\lambda x - x^{\alpha+1}} \left[\gamma(\alpha+1,\lambda \, x)\right]^{K-1}\, \de x \;.
\eea
This clearly shows that the distribution of the rescaled random variable $z = (E_0\,b)\,N^{\frac{1}{\alpha+1}}$ [see \eq(\ref{scale_E0})] converges to an $N$-independent form $F_k^{(\alpha)}(z)$ for large $N$, whose Laplace transform is given by $\int_0^\infty F_K^{(\alpha)}(z) \e^{-\lambda\, z} \, \de z = \langle \e^{-\lambda\,z}\rangle$. Therefore \eq(\ref{Laplace5}) demonstrates the result announced in \eq(\ref{expr_F_Laplace}).

\vspace*{0.5cm}

\noindent {\it Special cases $\alpha=0$}. In this case \eq(\ref{expr_F_Laplace}), using $\gamma(1,\lambda\,x) = 1 - \e^{-\lambda x}$, reduces to
\bea\label{special_a0}
\int_0^\infty F_K^{(0)}(z) \e^{-\lambda\, z} \, \de z = \frac{1}{\Gamma(K)\lambda^{K-1}} \int_0^\infty \de x\, \e^{-(\lambda+1)x} \left(1-\e^{-\lambda\,x} \right)^{K-1} = \frac{\Gamma(1+1/\lambda)}{\lambda^k \, \Gamma(k+1+1/\lambda)} \;.
\eea
Using the properties of the $\Gamma$ function, one can express the r.h.s.~of (\ref{special_a0}) as a simple product
\bea\label{special_a01}
\int_0^\infty F_K^{(0)}(z) \e^{-\lambda\, z} \, \de z = \prod_{m=1}^K \frac{1}{1 + m\,\lambda} \;.
\eea
To invert this Laplace transform, we note that the r.h.s.~has simple poles at $\lambda = - 1/m$ with $m = 0,1, \cdots, K$. Evaluating carefully the residues at these poles, we can invert this Laplace transform explicitly and get
\bea\label{special_a02}
F_K^{(0)}(z) = \sum_{n=1}^K (-1)^{K-n} \frac{n^{K-1}}{(K-n)! \, n!} \, \e^{-z/n} \;.
\eea
For instance,
\bea\label{a0_examples}
&&F_1^{(0)}(z) = \e^{-z} \\
&&F_2^{(0)}(z) = \e^{-z/2} - \e^{-z} \\
&&F_3^{(0)}(z) = \frac{3}{2} \e^{-z/3} - 2\,\e^{-z/2}+ \frac{1}{2}\,\e^{-z} \;.
\eea

\begin{figure}[bhtp]
    \centering
%    \subfigure[\label{fig:exponential} $p(\varepsilon) = e^{-\varepsilon}$]{
%        \includegraphics[scale=1]{exponential}
%    }
%    \subfigure[\label{fig:gaussian} $p(\varepsilon) = \frac{\sqrt{2}}{\sqrt{\pi}} e^{-\varepsilon^2/2} \Theta(0)$]{
%        \includegraphics[scale=1]{gaussian}
%    }
%
%    \subfigure[\label{fig:powerlaw} $p(\varepsilon) = \frac{2}{\varepsilon^3} \Theta(\varepsilon - 1)$]{
%        \includegraphics[scale=1]{powerlaw}
%    }
%    \subfigure[\label{fig:xexp} $p(\varepsilon) = \varepsilon e^{-\varepsilon}$]{
%        \includegraphics[scale=1]{xexp}
%    }
   \includegraphics[width = \linewidth]{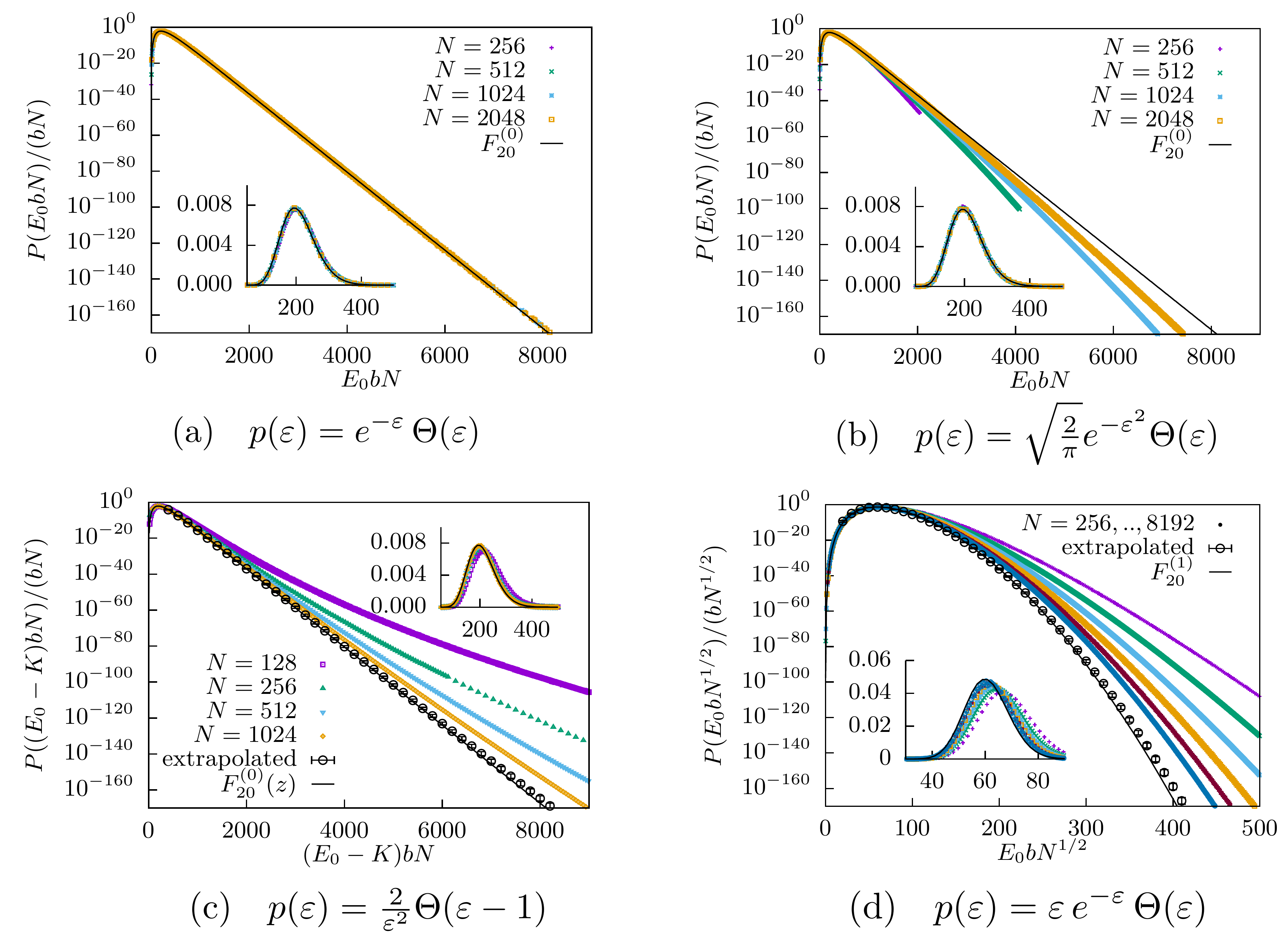}
    \caption{\label{fig:distributions}
        (colour online)
        Scaled distribution $P_{K,N}(E_0)$ for $K=20$, for different values of $N$ and for four different
        distribution $p(\veps)$. The insets show the behaviour near the peaks for the four different cases.
        (a) shows exponentially distributed
        $p(\varepsilon) = \e^{-\varepsilon}\, \Theta(\varepsilon)$ which corresponds to $\alpha = 0$ and $b=1$ [see \eq(\ref{main_result_form})]. The scaling function
        $F_{K=20}^{(0)}$ in \eq(\ref{special_a02}) matches very well the numerical data all the way up to the tails.
        (b) shows half-Gaussian distributed
        $p(\varepsilon) = \frac{\sqrt{2}}{\sqrt{\pi}} \e^{-\varepsilon^2/2}\, \Theta(\varepsilon)$ corresponding to $\alpha = 0$ and
        $b= p(0) = \frac{2}{\sqrt{2\pi}}$.
         (c) shows Pareto distributed energy levels
        $p(\varepsilon) = \frac{2}{\varepsilon^3} \Theta(\varepsilon - 1)$.
        After shifting $\varepsilon \to \varepsilon - 1$, \ie $E_0 \to E_0 - K$,
        this falls in the $\alpha = 0$ universality, with $b = 2$. In this case, using the finite-size extrapolation
        scheme shown in black circles (see text and \eq(\ref{fss}) with $\beta =1$), the data approach the theoretical scaling function
        $F^{(0)}_{20}(z)$.
        (d) shows energy levels distributed according to
        $p(\varepsilon) = \varepsilon \e^{-\varepsilon}\, \Theta(\veps)$. This corresponds to the
        $\alpha = 1$ universality class, with the scaling parameter $b = 1/\sqrt{2}$. Again, using the finite-size
        scaling form (see text and \eq(\ref{fss})) with $\beta = 1/2$, the extrapolated data (shown  in black circles) match well
        with the theoretical scaling function $F^{(1)}_{20}(z)$, obtained from the numerical Laplace inversion of \eq(\ref{expr_F_Laplace}), setting $K=20$ and $\alpha =1$.
    }
\end{figure}

\noindent{\it Numerical simulations.} Next, we verify our analytical predictions via numerical simulations. To test the prediction of the scaling behaviour in \eq(\ref{main_result_form}), as well as to test the universality of the associated scaling function $F^{(\alpha)}_K(z)$, we consider four different distributions for the energy levels:  (a) an exponential distribution
$p(\varepsilon) = \e^{-\varepsilon}\, \Theta(\veps)$, (b) an half-Gaussian
distribution $p(\varepsilon) =\sqrt{\frac{2}{\pi}}\, \e^{-\varepsilon^2}\, \Theta(\veps)$, (c)
a Pareto distribution $p(\varepsilon) = \frac{2}{\varepsilon^3} \Theta(\varepsilon - 1)$
and (d) $p(\varepsilon) = \varepsilon \e^{-\varepsilon}\, \Theta(\veps)$. The cases (a) and (b) clearly correspond to $\alpha = 0$. Hence we expect the scaling function to be given by $F_K^{(0)}(z)$ in \eq(\ref{special_a02}). The Pareto case (c), with support over $[1,+\infty)$, also corresponds to the $\alpha = 0$ case, as seen easily after a trivial shift $\varepsilon \to \varepsilon - 1$. Hence, in this case as well, we expect the scaling function to be given by $F_K^{(0)}(z)$. However, case (d) is different as it corresponds to $\alpha = 1$ and hence the scaling function should be given by $F_{K}^{(1)}(z)$. In \fig\ref{fig:distributions}, we compare the simulation results with the analytical predictions and find very good agreement. Note that in cases (a)-(c), the scaling function $F_K^{(0)}(z)$ has an explicit expression as in \eq(\ref{special_a02}). Hence, it is easy to compare directly the simulation results with this expression (as in \fig\ref{fig:distributions} (a)-(c)). However, for case $(d)$, where $\alpha = 1$, we do not have a simple explicit formula for $F_K^{(1)}(z)$, though we have explicitly given its Laplace transform in \eq(\ref{expr_F_Laplace}) with $\alpha =1$. Hence, to compare with the simulation results, we first needed to invert this Laplace transform using an arbitrary precision library \cite{mpmath}. This comparison is shown in \fig\ref{fig:distributions} (d).

To obtain the presented numerical results one has to
generate $N$ random numbers according to the
desired probability density $p(\varepsilon)$. This is done by a standard method, namely we first choose
a uniform random number $\eta \in [0,1]$ and then generate $\varepsilon$ using the formula, $\int_0^\varepsilon p(\varepsilon')\,\de\varepsilon' = \eta$. The exponential (a) and Pareto (c) case can be trivially obtained using this relation~\cite{recipes}. In the half-Gaussian case (b), the Gaussian random numbers can be generated using the Box-Muller method~\cite{recipes}. In the case (d), $p(\varepsilon) = \varepsilon \e^{-\varepsilon}$, the above relation reads
$\eta =\int_0^\varepsilon p(\varepsilon') \de\varepsilon' = 1-(1+\varepsilon) \e^{-\varepsilon}$, which can also be inverted
using the $-1$ branch of the Lambert W function~\cite{Corless1996lambertW}
$\varepsilon = -W_{-1}\left( \frac{\eta-1}{e} \right) - 1$. To evaluate the Lambert W function, we use the
GSL implementation~\cite{gsl}.

The sum $E_0$ in \eq(\ref{def_E0}) is completely determined by the
values $\vec \eta = (\eta_1,\ldots,\eta_N)$.
If one simply generates many times vectors $\vec \eta$ of independent
uniform random numbers and correspondingly obtained random numbers
$(\varepsilon_1,\ldots,\varepsilon_N)$, one will obtain only typical results
for $E_0$, \ie those
having a high enough probability. Here, we sample the distributions over a broad
range of the support,  also in the far tails, where the probabilities are extremely small.
For this purpose, we use a well tested importance
sampling scheme~\cite{Hartmann2002Sampling,Hartmann2014high}.
Here the vectors $\vec \eta$ are sampled using the Metropolis algorithm
including a bias of samples away from the main part of the
distribution. We use a bias $\e^{-E_0/T}$, where $T$ is a ``temperature'' parameter
which can be positive and negative and allows us to address different ranges of the distribution.
Since the bias is known, the Metropolis results can be corrected for the
bias to obtain the actual distribution.
This enables us to gather good statistics also in the far tails.

To be more concrete, we use a Markov chain $\vec \eta(t)=\vec \eta(0),\vec \eta(1), \ldots $
Every move $\vec \eta(t) \to \vec \eta(t+1)$ consists of changing
one  entry of $\vec \eta(t)$ leading to a trial $\vec\eta'$ (``local update'').
While the most simple method to change would
be the replacement of one uniform-distributed random number by a freshly drawn one,
as used in Ref. \cite{Hartmann2014high}, this will lead to
difficulties especially for small values $K$.
For the far tails, there will be a point
where all entries of $\vec\eta$ are almost one (or almost zero)
and almost every new proposal will be rejected, since it is improbable to draw
a random number very close to the previous one. Therefore we perform a slightly
more involved protocol, where instead of redrawing
we change an entry $\eta_i \to \eta_i + \xi \delta$,
where $\xi \in [-1, 1]$ is uniformly distributed and
$\delta \in \{ 10^{-i} | i \in \{0, 1, 2, 3, 4, 5\} \}$ with uniform probability $1/6$.
Thus $\delta$ determines the scale of the local change.
Changes resulting in an entry $\eta_i \not\in [0,1)$ are directly \emph{rejected},
\ie $\vec\eta(t+1)=\vec\eta(t)$. Note
that this protocol will still result in uniformly distributed random numbers $\eta_i$,
if every change in $[0,1)$ was \emph{accepted},
\ie $\vec\eta(t+1)=\vec\eta'$. Here
each proposed change is accepted instead with the Metropolis
acceptance ratio
\begin{align}
    p_\mathrm{acc} = \min\{1, \e^{-\Delta E_0 / T}\},
\end{align}
where $\Delta E_0$ is the change in energy caused by the proposed change,
and otherwise also rejected.

For any value of $T$, as usual for Monte Carlo simulations,
performing these Markov chains ``long enough''
and taking measurements, this results in a histogram $P_T(E_0)$ for each temperature,
which can be corrected for the bias using
\begin{align}
    P(E_0) = \e^{E_0 / T} Z(T) P_T(E_0).
\end{align}
The a-priori unknown normalization
 parameter $Z(T)$ can be obtained by enforcing continuity and normalization
of the whole distribution, which is obtained from performing simulations
for several values of $T$, including $T=\infty$, which corresponds
to simple sampling. We will not go into further details, since this
algorithm is well described in several other
publications \cite{Hartmann2002Sampling,Hartmann2014high,schawe2018avoiding}.

For the Pareto distributed case $p(\varepsilon) = \frac{2}{\varepsilon^3} \Theta(\varepsilon - 1)$,
we used instead of the aforementioned sampling with bias $\e^{E_0/T}$ a modified
\emph{Wang-Landau} sampling~\cite{Wang2001Efficient,Wang2001Determining,Schulz2003Avoiding,Belardinelli2007Fast,Belardinelli2007theoretical}
with multiple histograms
and subsequent entropic sampling~\cite{Lee1993Entropic,Dickman2011Complete}.
We used Wang Landau sampling for this case, since the temperatures are harder
to adjust, \ie for negative temperatures
it happens quickly that equilibration becomes impossible and the energy increases
constantly. This effect is already known to pose difficulties for the
aforementioned sampling with bias~\cite{Claussen2015Convex,schawe2017highdim}.

We set $K=20$ in \fig\ref{fig:distributions} and compare the distribution $P_{K=20,N}(E_0)$ for different values of $N$. We verify,
by a data collapse, the scaling form predicted in \eq(\ref{main_result_form}) and also compare the numerical scaling function to the analytical ones.
As mentioned earlier, for the $\alpha = 0$ case (corresponding to cases (a)-(c)), the analytical scaling function is given in \eq(\ref{special_a02}). For the $\alpha=1$ (corresponding to case (d)), we invert the Laplace transform in \eq(\ref{expr_F_Laplace}) for $K=20$ and $\alpha = 1$,
using an arbitrary precision library \cite{mpmath}.

\begin{figure}[t]
%    \centering
%    \subfigure[\label{fig:xexp_scaled_asymptote_left} left tail]{
%        \includegraphics[scale=1]{xexp_scaled_asymptote_left}
%    }
%    \subfigure[\label{fig:xexp_scaled_asymptote} right tail]{
%        \includegraphics[scale=1]{xexp_scaled_asymptote}
%    }
   \includegraphics[width = \linewidth]{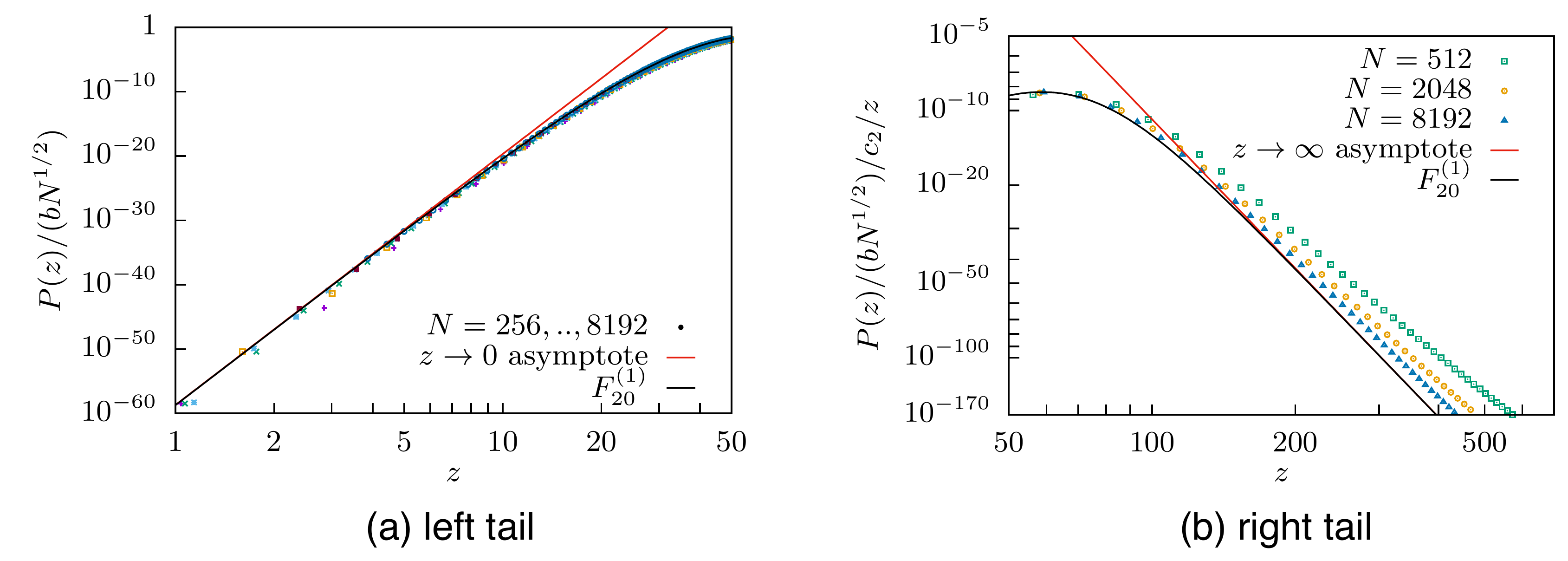}
    \caption{\label{fig:distributions_tail}
        (colour online)
        We consider the tails of the distribution $P_{K=20,N}(E_0)$ for the case $p(\varepsilon) = \varepsilon \e^{-\varepsilon}$, corresponding to $\alpha =1$. This scaling function in this case is given by $F_{K=20}^{(1)}(z)$. The asymptotic behaviors of $F_{K=20}^{(1)}(z)$ in \eq(\ref{F_asymptotics}), for $z \to 0$ (left tail, in (a)) and $z \to \infty$ (right tail in (b)) are compared to numerical simulations.
        The data have been plotted on a scale such that the two cases
        from \eq(\ref{F_asymptotics}) appear as straight lines. In (b), for clarity, only three different values of $N$ have been plotted.
  }
\end{figure}

While the exponential case fits very well to the analytic result
even for small values of $N$,
the other
cases show strong finite-size effects especially in the extreme right tail. Such
finite-size effects are known to occur frequently in the extreme statistics
of i.i.d.~random variables \cite{Racz}.
 As seen in \fig\ref{fig:distributions},
the discrepancy between the numerical and the analytical results is very small in the main region (\ie in the bulk).
In the tails, we need to use a finite-size ansatz to study the convergence of the numerical results as $N \to \infty$. For example, it
is natural to expect that the finite-size corrections to the leading scaling form in \eq(\ref{main_result_form}) is of the form
\begin{eqnarray}\label{fss}
P_{K,N}(E_0) =b \,N^{1/(1+\alpha)} \left [ F_K^{(\alpha)}(z) + N^{-\beta} G_K^{(\alpha)}(z) + N^{-2 \beta}H_K^{(\alpha)}(z) + \ldots  \right] \;,
\end{eqnarray} where $\beta = \min(1/(1+\alpha),1)$, $z = b\,
N^{1/(\alpha +1)} E_0$ is the scaling variable, and
$G_K^{(\alpha)}(z), H_K^{(\alpha)}(z)$ describe the finite-size
scaling of the correction terms. Thus for $\alpha=0$ one has $\beta =
1$, while for $\alpha = 1$, we have $\beta = 1/2$. For several values of $z$,
we extrapolate the data by fitting pointwise the
numerical data in \fig\ref{fig:distributions} as a function of $N$,
to obtain estimates for the asymptotes $F_K^{(\alpha)}(z)$. We treated the cases
(c) (corresponding
to $\alpha = 0$, and hence $\beta = 1$) and
(d) ($\alpha = 1$, $\beta = 1/2$), the extrapolated values are shown as symbols.
Furthermore, in  \fig\ref{fig:distributions_tail}, we
show the behaviour in the tails for the case (d), which
exhibits the strongest finize-size effects, such that the asymptotic
behaviour \eq(\ref{F_asymptotics}) is directly visible. It is apparent
that the convergence
for large values of $N$ is faster in the left tail $z \to 0$, while it is much
slower in the right tail $z \to \infty$.

\vspace*{0.3cm} {\noindent}{\it Conclusion.}  In this paper, we have
studied analytically and numerically
the full distribution of the ground-state energy
of $K$ non-interacting fermions in a disordered environment, modelled
by a Hamiltonian whose spectrum consists of $N$ i.i.d.~random energy
levels with distribution $p(\veps)$ (with $\veps \geq 0$), in the same
spirit as the ``Random Energy Model''. This ground state energy is the
sum of the smallest $K$ values drawn from a probability distribution
and therefore a generalization of the extrem-value statistics,
which corresponds to the case $K=1$. Thus our results should be of interest
also in a very general mathematical context.

We have shown that for each
fixed $K$, the distribution $P_{K,N}(E_0)$ of the ground-state energy
has a universal scaling form in the limit of large $N$ (see
\eq(\ref{main_result_form})). This universal distribution depends
only on $K$ and the exponent $\alpha$ characterizing the small $\veps$
behaviour of $p(\veps) \sim \veps^\alpha$. We derive an exact
expression for the Laplace transform of this scaling function in
\eq(\ref{expr_F_Laplace}). For generic $\alpha$, the asymptotic
behaviors of the scaling function are derived explicitly in
\eq(\ref{F_asymptotics}), while for the special case $\alpha = 0$,
the Laplace transform can be explicitly inverted, giving the full
scaling function in \eq(\ref{special_a02}). Numerically, while the
peak region of the distribution of $E_0$ can be easily estimated by
standard methods, estimating the tails of the distribution where the
probability is very small is hard and requires more sophisticated
techniques. In this paper, using an importance sampling algorithm, we
were able to estimate the tail probabilities (up to a precision as
small as $10^{-160}$) and thereby to verify the theoretical
predictions. Thus the main conclusion of our work is that, even though
the individual energy levels are independent random variables, the
ordering needed to compute the ground-state energy induces effective
correlations between the energy levels. These effective correlations
then lead, for the ground-state energy, to a whole new class of
universal scaling functions parameterised by $K$ and~$\alpha$.

In this work, we have modelled the single-particle energy levels of a quantum disordered system by i.i.d.~random variables, \`a la REM. This REM approximation for the energy levels is known to be valid for disordered Hamiltonians whose eigenstates are strongly localised in space \cite{MNS}. Thus we expect that the results presented in this paper for the universal distribution of the ground state energy
would apply to such strongly disordered quantum systems. It is then natural to ask what happens to the ground-state energy for Hamiltonians with weakly localised eigenstates. In some weakly localised systems, a description based on Random Matrix Theory (RMT) \cite{MNS} is a good approximation, where the energy levels (identified with the eigenvalues of a random matrix) are strongly correlated with mutual level repulsion. In this RMT context, several linear statistics of ordered eigenvalues have been recently introduced and studied for large $N$ under the name of truncated linear statistics (TLS) \cite{TLS1, TLS2}. The ground-state energy in \eq(\ref{def_E0}) or more generally the linear statistics as in \eq(\ref{Laplace4}) studied here are instances of TLS, but for i.i.d.~random variables. It would thus be interesting to see how the TLS, studied here for i.i.d.~variables, crosses over to the RMT case, as one goes from the strongly localised part of the spectrum of a disordered Hamiltonian to the weakly localised~part.

\begin{acknowledgments}
    This work was supported by the German Science Foundation (DFG) through
    the grant HA 3169/8-1. HS and AKH thank the LPTMS for hospitality and
    financial support during one and two-month visits, respectively, where
    this project was conceived.
    The simulations were performed at the cluster of the GWDG G\"ottingen and
    the HPC cluster CARL, located at the University of Oldenburg (Germany) and
    funded by the DFG through its Major Research Instrumentation Programme
    (INST 184/157-1 FUGG) and the Ministry of Science and Culture (MWK) of the
    Lower Saxony State. SM and GS acknowledge support by ANR grant ANR-17-CE30-0027-01 RaMaTraF.
\end{acknowledgments}

%\bibliography{lit}

\end{document}